\UseRawInputEncoding
\documentclass[a4paper,11pt]{article}
\usepackage{pos}

\usepackage[]{graphicx}
\usepackage{float}

\newcommand{\um }{$\mu$m}
\newcommand{\cmq}{cm{$^{-3}$}}

\newcommand{\Msol}{M{$_{\odot}$}}

\newcommand{\Lsol}{L{$_{\odot}$}}

\newcommand{\Feii}{Fe~{\sc ii}}
\newcommand{\Sii}{S~{\sc ii}}
\newcommand{\Hii}{H{\sc ii}}
\newcommand{\Oi}{O{\sc i}}
\newcommand{\Oii}{O{\sc ii}}
\newcommand{\Oiii}{O{\sc iii}}

\newcommand{\Ha}{H${\alpha}$}
\newcommand{\Nii}{N{\sc ii}}

\newcommand{\Htwo}{H{$_2$}}

\newcommand{\mm}{$\mu$m}

\newcommand{\kms}{km~s$^{-1}$}

\title{Jets, Outflows, and Explosions in Massive Star Formation }

\author*[a]{John  Bally}

\affiliation[a]{Center for Astrophysics and Space Astronomy (CASA) \\
                    Department of Astrophysical and Planetary Sciences (APS) \\
                    University of Colorado, Boulder \\
                    2000 Colorado Blvd., Boulder,  USA  }

\abstract{
Multispectral studies of nearby,  forming stars provide insights into all classes of accreting systems.  Objects which have magnetic fields, spin, and accrete produce jets and collimated outflows.    Jets are seen in systems ranging from brown dwarf stars to supermassive black holes. Outflow speeds are typically a few times the escape speed from the launch region - 100s of \kms\ for young stars to nearly the speed of light for black-holes.  Because many young stellar objects (YSOs) are nearby, we can see outflow evolution and measure proper motions on times scales of years.   Because the shocks in YSO outflows emit in atoms, ions, and molecules in addition to the continuum, many physical properties such as temperatures, densities, and velocities can be measured.   Momenta and kinetic energies can be computed.   YSO outflows are a major source of feedback in the self-regulation of star formation.  The lessons learned can be applied to much more distant and energetic cosmic sources such as AGN and galactic nuclear super winds - systems in which evolution occurs on time-scales of hundreds to millions of years.   Some dense star-forming regions produce powerful explosions.   The nearest massive star-forming region, Orion OMC1, powered a $\sim 10^{48}$ erg explosion about  550 years ago (that is when the light from the event would have reached the Solar System).  The OMC1 explosion was likely powered by an N-body interaction which resulted in the formation of a compact, AU-scale binary or resulted in a protostellar merger.   The binary or merger remnant, the $\sim$15 \Msol\ object known as radio source  I  (Src I)  was ejected from the core with a speed of $\sim$10 \kms\ along with two other stars.   The $\sim$10~\Msol\ BN object was ejected with $\sim$30~\kms\ and a $\sim$3~\Msol\ star was ejected with $\sim$55~\kms . 
}

\FullConference{
Multifrequency Behaviour of High Energy Cosmic Sources XIV (MULTIF2023)\\
12-17 June 2023\\
Palermo, Italy\\}

\begin{document}
\maketitle

\section{Introduction}

Accreting astrophysical systems with rotation and magnetic fields tend to exhibit bipolar outflows or collimated jets.  Examples include young stellar objects (YSOs), symbiotic stars, post main-sequence objects such as proto-planetary nebulae,  planetary nebulae, some supernovae, accreting neutron stars such as SS433, micro-quasars, active galactic nuclei,  and quasars.       Their morphologies consist of  collimated, oppositely-directed beams near the source which break into  chains of knots and bow shocks farther out.    The jets create cocoons of ejecta surrounded by material swept-up  from the environment and entrained into bipolar flows.   The jet speeds tend to be a few times the circular orbit speed where  the outflow is launched - a few hundred \kms\ for protostars and near the speed of  light for neutron stars and black holes.   Dimensionless parameters such as the ejecta speed divided by the orbital speed at the launch point, the degree of collimation, morphology, and variability in mass-loss-rate, ejection velocity,  ejection direction, and degree of collimation are similar in all classes of accretion-powered outflows.

\begin{figure}[!ht]
\begin{center} 
\includegraphics[width=6in]{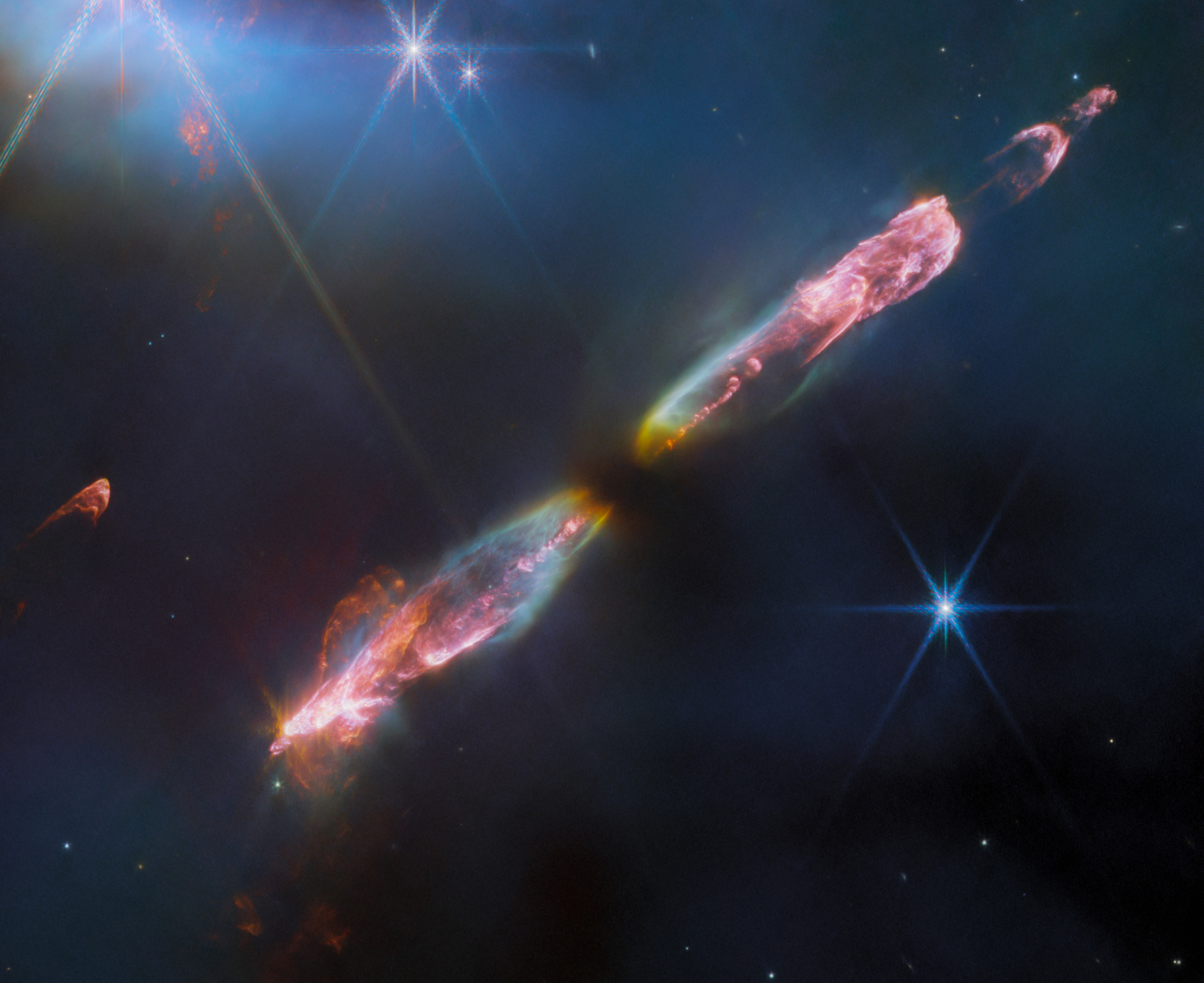}
\end{center}
\caption{A JWST image showing the HH~211 jet and bipolar outflow emerging from a low-mass protostar, hidden by the dusty core in the center of the image.   The image shows shock-excited molecular hydrogen emitting at a wavelength of 4.69~\mm\ (orange-red) and a combination of carbon-monoxide (CO) emission in the v=1-0 vibrational band at 4.6~\mm \ and continuum (cyan-blue).      HH~211 is located in the Perseus Molecular cloud at a distance of $\sim$300 pc.    See Ray et al. (2023) for details.
             }
\label{fig1}
\end{figure}

Young stellar objects (YSOs)  constitute the nearest class of accretion-powered jets 
and outflows. (Bally 2016; Reipurth and Bally 2001).    
Because the ejecta and entrained material consists of a mixture of 
molecules, atoms, and ions (instead of just fully ionized plasma) 
which shine in emission lines, densities, temperatures,  
and radial velocities can be measured.   These measurements allow the momentum and
kinetic energy of the outflow to be determined.   Because of their proximity, 
changes and proper motions can be seen on HST, JWST, and adaptive-optics-assisted
images taken a few years apart.   
YSO outflow structure and motions  provide a fossil record of the 
mass-loss and  accretion histories of the source stars.  YSO outflow shocks
trace the momentum and energy injected into the ISM by protostellar 
feedback.  The high-resolution investigation of protostellar 
jets and their sources lead to insights into the physics of many classes 
of accretion-powered outflows.   

YSO outflows can be detected from visual to radio wavelengths.  Some of the fastest
outflows also emit  X-rays.   At visual wavelengths, 
the shock-excited nebulae powered
by outflows from young stars are called Herbig-Haro objects (Reipurth \& Bally 2001).     
In the near-infrared, these outflows power Molecular Hydrogen
Objects (MHOs) dominated by the emission lines of molecular hydrogen (\Htwo ).  
At mm and sub-mm wavelengths YSO outflows are traced by the
rotational  transitions of various molecules, especially carbon monoxide (CO)
and silicon monoxide (SiO).   Some jets and outflows are also visible in the radio
continuum.

Figure 1 shows an image of the bipolar jet and outflow,  HH~211, emerging from a
low-mass, extremely young,  Class~0 protostar embedded in a cloud core 
at the eastern end of the Perseus molecular cloud (D$\approx$300 pc).     
The protostar has a mass  of only 0.08~\Msol\ 
but is surrounded by a 0.2~\Msol\ envelope from which it continues to accrete.  
The protostar drives bipolar molecular jets rendered visible by shock waves
which excite transitions of \Htwo\ and CO.   Figure 1 shows the 4.69~\mm\ v=0-0 S(9) 
transition in molecular hydrogen and the 
v=1-0 vibrational band of carbon monoxide (CO) at 4.6~\mm .   The jet has 
a mean speed of $\sim$100~\kms .  Velocity variations ranging from 80 to 120 \kms\ 
result in collisions where faster fluid overtakes slower ejecta.  These internal working
surfaces excite \Htwo\ into emission.     Over time, the jet has inflated a cocoon 
which has swept-up the ambient medium into elongated, twin bubbles visible in CO 
and scattered light.  The outflow is also visible in the  pure rotational transitions  of 
CO and SiO at millimeter-wavelengths.

\section{Overview of Star Formation}

\subsection{GMCs and the Ecology of Star Formation}    

Most stars form in molecular clouds experiencing gravitational collapse and fragmentation
(for a review, see Bergin \& Tafalla 2007). 
Most of the $\rm M_{tot}(H_2)  \sim 2$ to $5 \times 10^9$~\Msol\ of molecular gas in the 
Milky Way is located in Giant Molecular Clouds (GMCs) which populate the Galactic 
disk (Murray et al. 2011; Dame et al 2001).  GMCs have typical masses of $10^4$ to
$10^6$~\Msol\ and sizes of 10s of parsecs.  GMCs are cold with temperatures
of 5 to 20 K and  sound speeds  $c_s \sim$0.1 to 0.25 \kms .  
The observed molecular line widths are typically a few \kms ,  implying that GMCs are supported 
by supersonic turbulence with Mach numbers ranging from 2 to over 20.   Magnetic fields tend to 
be in equipartition with  turbulent motions, implying field strengths  of 
10s to 100s of  micro-guass (Crutcher 2012).     GMCs in the Galactic disk have 
mean densities of $\rm N(H_2) \sim 10^2 - 10^3$~\cmq .   

\begin{figure}[!ht]
\begin{center} 
\includegraphics[width=6in]{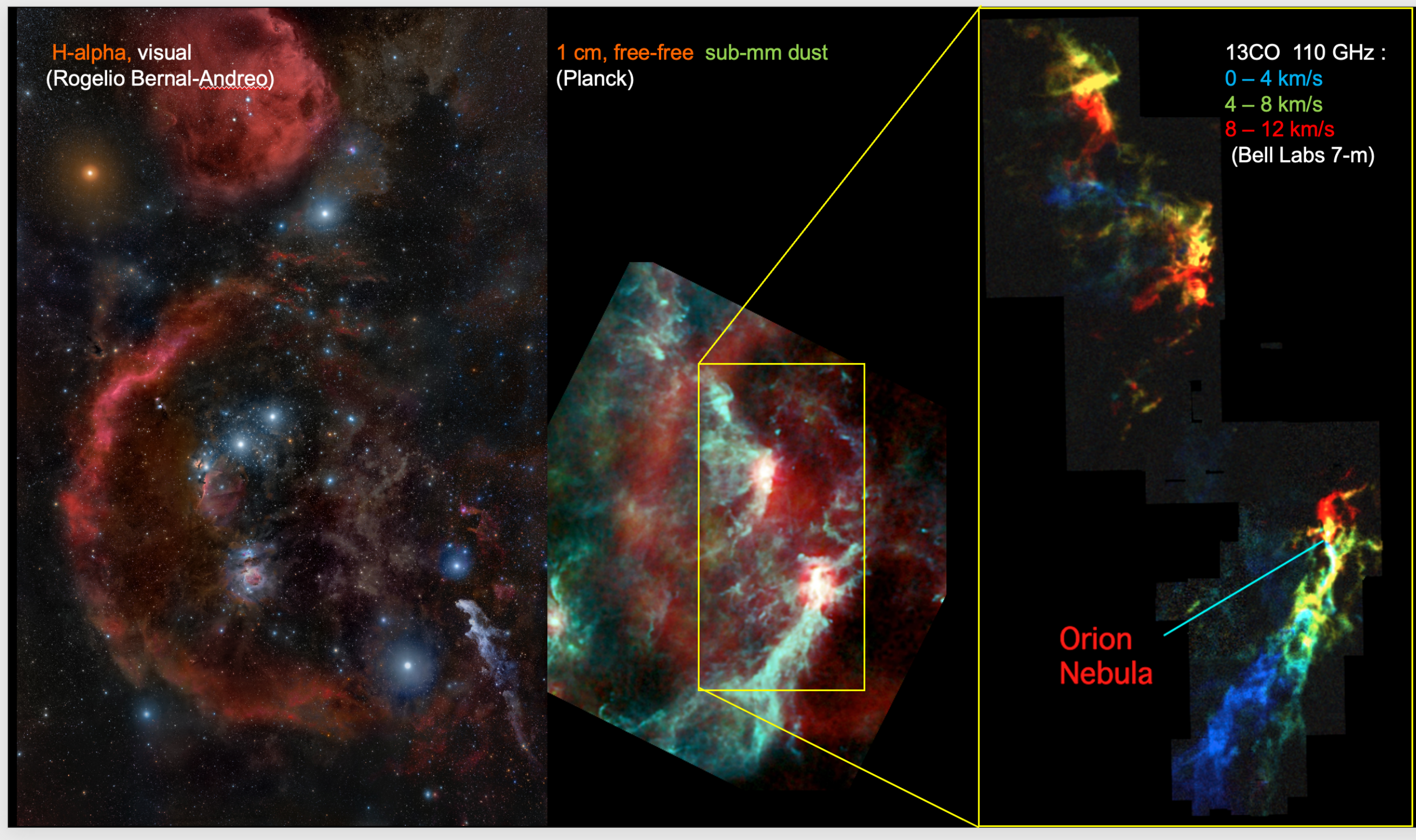}
\end{center}
\caption{
{\bf (Left:)} A wide-field, visual wavelength image showing the entire constellation of Orion.    The prominent red arc is Barnard's Loop, thought to be a shell swept-up by the most recent supernova in the Orion OB1 association.   
{\bf (Middle:)}  The lower and central part of Orion in an image obtained by the Planck satellite shown at the same scale as the visual wavelength image on the left.  Red shows thermal free-free emission at 30 GHz (1 cm).  Barnard's Loop is clearly visible.   Green shows emission at 545 GHz mostly tracing warm $\sim$20 K dust.  The yellow box outlines the field shown in the right panel and contains the $\sim$50,000 \Msol\ Orion A and Orion B molecular clouds.
{\bf (Right:)}    A 110 GHz (2.6 mm wavelength) image showing $^{13}$CO emission from the Orion A and Orion B molecular clouds.   Colors indicate the radial velocity of the emitting gas in the LSR frame revealed by the Doppler shift of the $^{13}$CO emission line.     Blue shows $\rm V_{LSR}$ = 0 to 4 \kms ;   green shows 
$\rm V_{LSR}$ = 4 to 8 \kms ;   red shows $\rm V_{LSR}$ = 8 to 12 \kms .    Taken with the AT\&T Bell Labs 7-meter antenna (Bally et al. 1987).
             }
\label{fig2}
\end{figure}

\begin{figure}[!ht]
\begin{center} 
\includegraphics[width=6in]{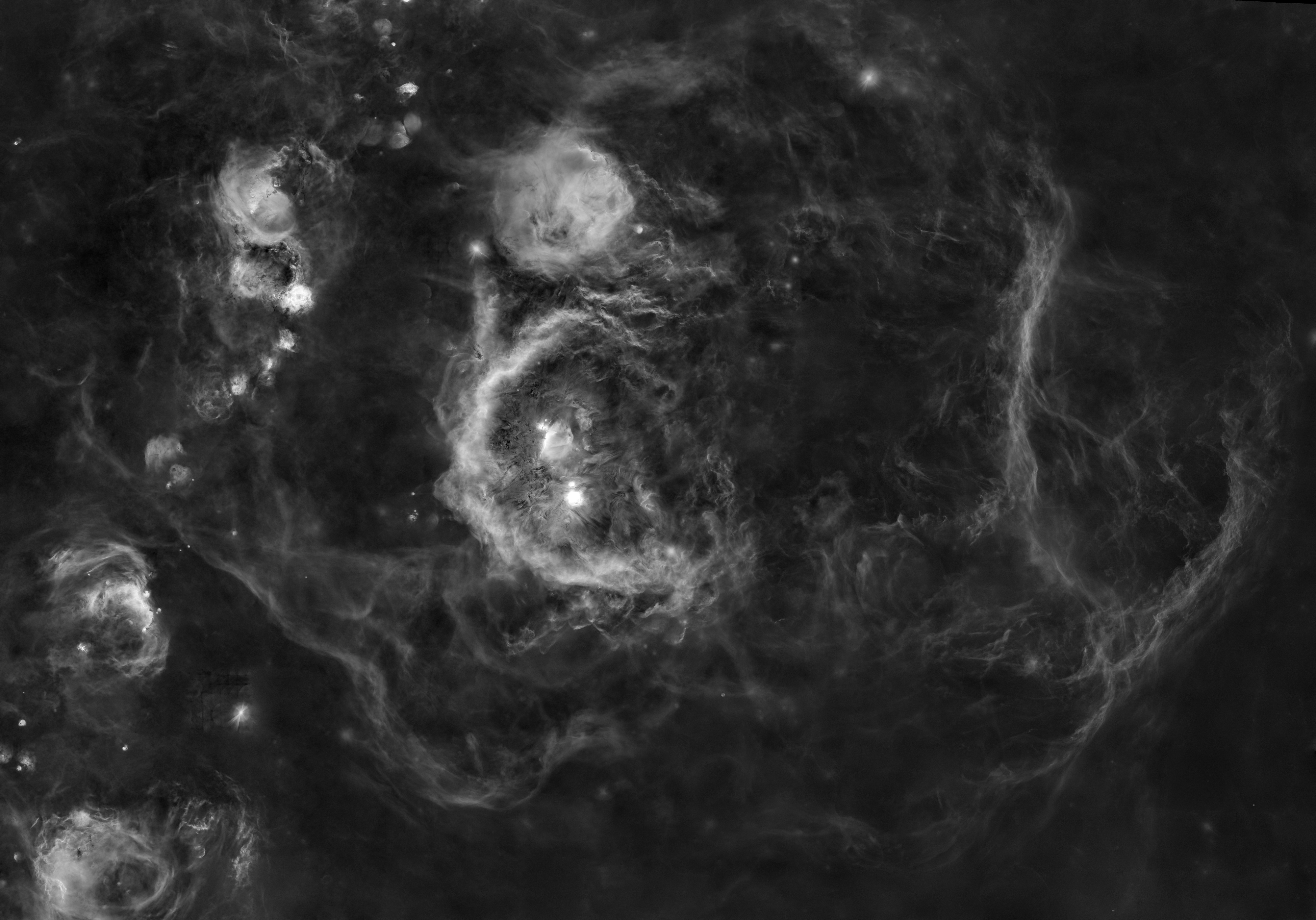}
\end{center}
\caption{A wide field-view showing the $\sim$300 pc long Orion-Eridanus superbubble in \Ha\ which was created by the combined impact of ionizing radiation, radiation pressure, stellar winds, and multiple supernova explosions.   Tens of thousands of stars formed in GMCs located near the center of the superbubble over the last 15 million years.   The bright arc left of center, Barnard's Loop, is thought to be a shell swept up by a supernova which exploded  $\sim 3 \times 10^5$ years ago.  The smaller bubbles near the left
edge trace \Hii\ regions located in the background along the Galactic plane in the anti-center direction. 
Most stars have been removed from this image.  However, residuals from the brightest stars such Betelgeuse and Rigel in Orion are visible.  Sirius is near the lower-left corner.  Aldebaran in the Hyades is visible right of center near the top.  This image was obtained as part of the MDW (Mittelman, DiCicco, Walker) \Ha\ survey  by Dennis DiCicco and Sean Walker.  
             }
\label{fig3}
\end{figure}

GMCs located in the Central Molecular Zone (CMZ), the region of the Galaxy at galactocentric radii less 
than $\sim$500 pc,  are warmer (T$_{gas} \sim$50 to 200 K), more turbulent 
($\rm \Delta V \sim$ 10 to 50 \kms ), and orders of magnitude denser ($\rm N(H_2) > 10^4$~\cmq )
than clouds  in the Galactic disk located more than a few kpc from the Galactic center. 
CMZ molecular clouds exist in a high pressure, high-shear environment as they orbit
the Galactic center.   Their physical conditions may be similar to star forming environments
in the distant, high-redshift Universe.

The Orion star forming complex (Figures 2 and 3), located at a distance of $\sim$400 pc, is the nearest region 
supporting on-going massive star formation (Gro{\ss}schedl et al. 2021).    Two GMCs in Orion each have masses 
of about $\rm 5 \times 10^4$~\Msol\ (Figures 3 \& 4).  The southern cloud, called Orion A, resembles a giant, 20~pc-long comet pointing towards the oldest stars in the Orion OB association and the center of the $\sim$300 pc diameter "superbubble" created by the action of massive stars over the last $\sim$15 Myr (Figure 3).    The northern  portion of the Orion A cloud, the so-called `Integral-Shaped-Filament' (ISF) located at the head of the comet,  is the densest, most compressed part of Orion A.    The Orion Nebula is located directly in-front of the upper portion of the Orion A cloud in the middle of the ISF.    Within the last few million years, dense clumps in the ISF  spawned at least 2,000 stars, creating the Orion Nebula Cluster.   A half dozen OB stars in this cluster  are responsible for ionizing the Orion Nebula.  

The Orion B cloud is located east of the superbubble center.  Its western edge is where  current star formation is occurring.   The \Hii\ region NGC~2024, the soft-UV  created bubble NGC~2023, and the famous Horsehead Nebula are located here. The IC~2118 cloud,  illuminated by the star Rigel, is located in the lower right portion of the left panel in Figure 2.    Like the Orion A cloud, IC~2118 is  a cometary cloud whose axis points towards the center of expansion of the Orion-Eridanus superbubble.  

Over the last 10 to 15 million years,  
Orion has given birth to tens of thousands of stars, including dozens of OB stars with masses larger 
than $\sim$8~\Msol .  At least a dozen massive stars have already exploded as supernovae.  The
collective impact of ionizing radiation, stellar winds, and supernova explosions have carved out a $\sim$300 pc diameter superbubble in the interstellar medium (Figure 3; Ochsendorf et al. 2015).   The bubble is surrounded by a swept-up,  $10^5$~\Msol\  shell of atomic and ionized hydrogen.  Because the Orion region is located about 150~parsecs below the Galactic plane, the bubble preferentially expanded toward high Galactic latitudes in the constellation of Eridanus.   Thus, this bubble is called the Orion-Eridanus superbubble.    The most recent supernova explosion is thought to have occurred about 300,000 years ago and may be  responsible for the arc of \Ha\ emission located along the eastern side of the Orion constellation  known as Barnard's Loop  (Foley et al. 2023).  

\subsection{Gravitational Collapse, Fragmentation, Star Formation, and Feedback}

In the absence of forces opposing the tendency of clouds to collapse, the free-fall collapse
time of a GMC would be $\rm \tau_{ff} \sim (G~\rho)^{-1/2}$ $\rm  \sim6 \times 10^6~ n^{-1/2}_{100}(H_2)$~years
where $\rm \rho = \mu n(H_2) m_H$ and $\rm n_{100}(H_2)$ is the molecular hydrogen density 
in units of 100 \cmq .   This would imply a Galactic star formation rate, 
$\rm SFR \sim M_{tot}(H_2) / \tau_{ff} $
= 300 to 900 \Msol yr$^{-1}$, orders-of-magnitude higher than the 
{\bf observed Galactic star formation rate of only 
$\approx$2 \Msol yr$^{-1}$!}

Apparently, feedback from young stars, turbulence, and magnetic fields effectively oppose gravitational 
collapse and limit the rate at which GMCs form stars.    The observed star formation rate implies that the
depletion time-scale of molecular gas in the Galaxy is about 1 to 2 billion years.  However, because of feedback, individual GMCs have lifetimes of less than 10 to 20 million years.   Most GMCs are destroyed by  ionization by extreme ultraviolet  (EUV) radiation,  winds, and SN explosions by the time $\sim$few  to 15\% of their mass has been converted into stars.     Some fragments of the parent GMCs  may survive and be accelerated away by the rocket effect (photo-ablation of plasma by EUV), winds, and blast waves.   However, most of the gas will be dissociated, ionized and dispersed.  Eventually, as the massive OB stars which formed from the GMC die within $\sim$40 Myrs of birth (the lifetime of an 8~\Msol\ star,   the least massive star which emits significant amounts of hydrogen-ionizing,  EUV photons with $\lambda <$ 912 \AA ,  and the least massive to explode as a supernova), the plasma recombines to form neutral HI clouds.  Eventually molecules reform.  Compression resulting from cloud-cloud collisions or the passage of a shock such as a spiral density wave can trigger cooling and gravitational instabilities, leading to the formation of new GMCs from the debris of the old.    

In the Solar vicinity, the time-scale for the cycling of atoms from a GMC to the hot interstellar medium (ISM) and back into a GMC is approximately $\sim$100 Myrs.   I call this cycling the "Galactic ecology".   Typically, only  $\sim$5\% of the gas in a GMC is locked into stars during any one cycle.   Most of these stars are long-lived, low-mass stars which survive for nearly the age of the Universe (but stars with masses less than 0.8~\Msol\ live longer than 14 billion years).  Thus, they remove their atoms from the Galactic ecology.    It is only the short-lived, luminous, high-mass stars which return most their matter to the ISM within $\sim$100 Myr of birth.   Some of this gas is converted into heavy elements by nuclear reactions in massive star cores,  increasing the metallicity of the Galaxy.    The remaining $\sim$95\% of the gas in a GMC continues to participate in the Galactic ecology cycle.    Any given atom has a $\sim$5\% change of getting locked-up in a star during one cycle.   But after 10 to 20 cycles (1 to 2 Gyr), chances are that this atom will end-up in a star.   In the absence of infall of fresh gas from outside the Galaxy, star formation will deplete the ISM within 1 to 2 Gyr.

Star formation in magnetized and turbulent clouds can occur because of ambipolar diffusion and/or 
the decay of turbulence.  In some circumstances,  star formation can be triggered by cloud-cloud
collisions,  external compression caused by passage of a spiral arm,  or the impact of radiation or
a blast wave created by an expanding \Hii\ region, stellar wind bubble, supernova remnant, or
superbubble.

When gravitational collapse overcomes turbulence and magnetic pressure in a GMC, the result is the
formation of over-dense clumps.  The collapse scale is given by the generalized Jeans length and 
Jeans mass.  The Jeans length is 
$\rm R_{J} \approx c_{s,eff} (G \rho)^{-1/2}$ where
$\rm c_{s,eff} \approx (c_s^2 + \Delta V^2_{turb} + V^2_{Alfven})^{1/2}$ is the `effective' sound speed which accounts for the effects of turbulence and magnetic fields.  Here, 
$\rm c_s$ is the sound speed, $\rm  \Delta V_{turb}$ is the turbulent velocity dispersion, and
$\rm V_{Alfven}$ is the Alfv\'en speed.    The Jeans mass is 
$\rm M_J \approx (4 \pi / 3) \rho R^3_J$.  For a mean GMC density $\rm m(H_2)$=100~\cmq\ and
$\rm  \Delta V_{turb}$=1~\kms , 
$\rm R_{J} \approx$ 6 pc and $\rm M_J \approx$ 6,000~\Msol .   
Observations show that star forming clumps have densities of $\rm n(H_2) > 10^4$~\cmq ,
sizes less than a parsec, and masses ranging from 100 to $\rm 10^4$~\Msol .

The Jeans length decreases with increasing density as $\rho ^{-1/2}$ and  
the Jeans mass decreases as  $\rho ^{-3/2}$.  
Thus,  collapsing clumps tend to fragment into clusters of denser cores.   
The fragmentation cascade radiates away the gravitational potential energy released during 
collapse  as long as the cores remain optically thin to the characteristic 
sub-mm radiation produced  by dust with a temperature of 10s of Kelvin 
and a variety of spectral lines of species such as CO  which can be collisionally excited in the cool gas.
Large clump masses and fragmentation  imply that
most stars form in clusters containing hundreds to thousands of stars 
in a region typically no larger that a parsec in diameter.  

The highest densities tend to occur at the clump and core centers.  
Thus, the collapse tends to proceed in an inside-out
manner.   When the column density grows to about  
$\rm N(H_2) >10^{25}$~cm$^{-2} \approx$ 0.1~g~cm$^{-2}$, dust
 becomes optically thick to its own radiation.    The cores start to warm. 
When central temperatures reach a few thousand K, \Htwo\ dissociates into atoms.  As
the core reaches many thousands of Kelvin,  hydrogen atoms are collisionally ionized;  a protostar
is born with an initial mass of 0.01 to 0.1 \Msol .   

\subsection{Most Stars form in Short-Lived, Transient Clusters}

Observations show that almost all stars are born in clusters containing tens to thousands of
stars  (Lada \& Lada 2003). Isolated single
star formation is rare.   In most  clumps, feedback  by jets, radiation, winds, and explosions results in  
low star formation efficiency.  The  star formation efficiency (SFE) is defined as the total mass 
of all stars produced divided by the initial mass of the star forming clump.  Low SFE implies 
that the forming clusters will be short lived.  Stars in such a cluster
inherit the velocity dispersion of the parent clump, roughly given by 
$\rm \Delta V \approx (G M_{clump} / R_{clump})^{1/2}$ $\sim$5 to 10~\kms\ which will be similar to the widths of spectral
lines emitted by the clump's molecular gas.  If 70\% or more of the mass of the clump
is expelled by various feedback processes, the gravity of the stellar cluster itseff will be insufficient to 
bind the stars to the cluster.   Once most gas is expelled by feedback,  the new-born stars will drift away and
the cluster will dissolve  on a few crossing times, 
$\rm t_{cross} \sim R_{cluster} / \Delta V$ $\sim 2$ to $10 \times 10^5$~years.

How do bound stellar clusters form which survive for hundreds of crossing-times or longer?  
If the SFE is greater than $\sim$30\%, numerical models show that
clusters can survive for many crossing times to become  open clusters such as h and $\chi$-Persei
or the Pleiades.    In this case, cluster survival time is determined by 
N-body dynamics and stellar evolution.  Massive stars (M$>$8~\Msol) die within 
40 Myr (for a $\sim$8~\Msol\ star) to $\sim$3 Myr (for a $\sim$100~\Msol\ star) after birth.     
Open clusters tend to evolve toward equipartition of stellar kinetic energy. This implies that
massive stars tend to sink towards the cluster core while lower mass stars tend to be ejected or move onto
orbits taking them away from the core.     In this scenario, tightly bound stellar systems which can survive
for a Hubble time (such as globular clusters) require extremely high SFE, approaching or exceeding 90\%.
Such conditions may be met in super-dense and compact GMCs containing more than $10^6$~\Msol\ within a
radius of only a few parsecs.  Today, such clouds are only found in galactic centers and in galaxies that have experienced recent mergers.  However, such clouds may have been more abundant in the early Universe when a larger fraction
of atoms were in the ISM and fewer in stars.

Most cores fragment further to produce multiple stars, often in non-hierarchical 
configurations which are subject to rapid decay and re-configuration into stable hierarchical groups
(Reipurth et al. 2010; Reipurth \& Mikkola 2012, 2015).   Dynamics of such multi-body  systems leads
to the decay of most multiples to generate the field-star multiplicity statistics.   The decay of unstable 
multiples can eject forming stars from their parent cores, thereby setting their final masses.   Although over 80\% of
stars form in multiple systems,  in the field, most low-mass stars are single.  However,  most massive stars
are multiple (Zinnecker \& Yorke 2007).   

In a cluster, two processes likely dominate the termination of accretion:  Feedback from other stars and N-body dynamics.    As stars form and grow in mass they generate a sequence of ever-more powerful feedback impacts.   Accretion, spin, and magnetic fields conspire to produce bipolar jets.    Accretion luminosity and eventually nuclear fusion  result in the production of UV radiation.  Soft-UV dissociates and heats molecules.  Harder EUV radiation with wavelengths below 912 \AA\ ionizes hydrogen.
Stars between $\sim$10 and 20 \Msol\ which have stopped accreting at high rates, and all stars above $\sim$20~\Msol\  produce mostly ionizing radiation which creates \Hii\ regions.   Such stars also produce powerful line-driven winds with 
speeds of $\sim10^3$~\kms\ and mass loss rates of order $10^{-8}$ to $10^{-6}$~\Msol~year$^{-1}$.  
During their post-main sequence evolution these massive stars drive even more powerful winds. Then they explode as supernovae.    Remnant neutron stars and black holes can inject even more energy in the form of pulsar-wind nebulae and  relativistic jets if they survive in a mass-transfer binary.
I call the sequence of ever more powerful feedback impacts (jets, soft-UV, ionizing UV, winds, supernovae, etc.) the "feedback ladder".    Feedback drives turbulence in the surviving clump and GMC.   A clump can be destroyed by feedback on a time scale of a few Myr.   On a time-scale of $\sim$5 to 20 Myr, the entire GMC can be destroyed.    

N-body dynamics in a forming cluster can also terminate accretion.   Gas in  cluster-forming clumps makes for a viscous medium.   Bondi-Hoyle accretion and dynamic friction (the gravitational attraction of a concentration of gas created by gravitational focusing behind a moving mass)  facilitates orbit decay, enabling the migration of cores and YSOs toward the center of the gravitational potential well.     The high density of YSOs and cores enables interactions;   cores with cores, stars with cores, and stars with stars.  Interactions tend to strip-off gas beyond the gravitational radius, $\rm r_G \approx GM / V^2$ where M is the mass of a given star or core and V is the speed of the interaction at closest approach.    An ejected protostar can only retain material within $\rm r_G$.  Thus, dynamical ejection from a core or clump sets the stellar mass.  In a forming cluster, N-body interactions may play an important role in establishing the Initial Mass Function (IMF) of stars.

\subsection{Spin, Disks, Magnetic Fields, and YSO growth}
  
In a turbulent GMC,  collapse and fragmentation tends to
produce clumps and cores which inherit some angular momentum from the parent GMC or clump.  Even in the absence of spin in the parent GMC, tidal forces between passing cores can convert linear motion into angular momentum and  spin.   The collapse of  cores with angular momentum tends to produce  flattened, rotating structures.   While low angular momentum material falls in close to the center, 
higher angular momentum matter will fall into the structure farther from the center.  Dissipation of motions parallel to the net angular momentum vector will lead to the formation of a disk.   Shear within the disk will tend to amplify entrained magnetic fields dragged in by the accretion flow.  

Protostars continue to grow by accretion from the surrounding, rotationally supported disk as long as there is supply of disk material and there is viscosity (dissipation) within the disk.   For quiescent cores with only sub-sonic turbulence,  the accretion rate onto the core center is roughly given by $\rm dM/dt \approx c^3_s / G$ $\sim 1.3 \times 10^{-6}$~\Msol ~yr$^{-1}$ for $\rm c_s$=0.17 \kms\ correponding to a gas temperature of 10 K.    This formula is derived  for a highly idealized, non rotating, isothermal core with a centrally condensed $\rm r^{-2}$ density profile.   Nonetheless, it provides a guide for estimating the formation time-scales for stars.   Thus, a 1 \Msol\ star is expected to take about one million years to accumulate its mass.    Observations show that stars with roughly 1 Solar mass actually takes a few Myr to complete their accretion.  This is still much shorter than the $\sim$20 to 30 Myr that it takes a Sun-like star to reach the main sequence.    

Note that the accretion luminosity of a  star is given by 
$\rm L_{acc} \approx G M (dM/dt) / r_{YSO}$ where M is the mass of the forming star and $\rm r_{YSO}$ is its radius.   For 
a YSO with a radius $\rm r_{YSO}=10^{11}$ to $\rm 10^{12}$~cm and a mass M = 1~\Msol ,  $ \rm L_{acc} \approx $28 to 2.8~\Lsol\  for the above accretion rate.    Thus, rapidly accreting low-mass stars tend to be over-luminous compared to their main-sequence luminosities. 

 After the initial collapse of a core, further stellar growth occurs by accretion from a circumstellar disk.  In this case, disk viscosity limits the accretion rate.   Magnetic fields in the disk can drive MHD turbulence which may be responsible for the viscosity that  transports angular momentum outward and mass inward.    Observations show that accretion tends to be episodic and variable in rate.     Accretion bursts  are followed by long periods of  quiescence.    These variations may be caused by substructure in the disk (spiral arms, clumps, etc) or various instabilities which result in large variations in viscosity.     During accretion bursts,  much of the inner disk may drain onto the star.  It may take  many disk  orbit time scales to replenish the inner disk.     Such variations in accretion rates are responsible for large variations in the luminosities of forming stars.   Because of the intimate connection between accretion and mass-loss, accretion-rate variability is also responsible for variations in the mass-loss rates and ejection speeds of outflows.   Such variations are responsible for the complex structures observed in the outflow ejecta.     
 
As stars accumulate mass, their luminosities grow.  Increasing luminosity results in warmer surrounding gas  and larger sounds speed.  This results in higher accretion rates.  Observations show that massive protostars - those with masses larger than 10 \Msol\  - accrete at rates ranging from $\rm 10^{-5}$ to over $\rm 10^{-3}$ \Msol~year$^{-1}$.    As mass increases, the time to reach the main sequence shrinks.    Unlike low-mass (Sun-like) stars, massive stars lack an observable pre-main sequence phase.  At  high accretion rates, models show that the photospheres of stars with masses between 10 to 20 \Msol\ become bloated and cool.  As long as the high accretion rate is maintained, they resemble red giants or super-giants.    Once accretion stops, or if the stars reach a mass above $\sim$20~\Msol , they settle onto-the main-sequence on a Kelvin-Helmholtz time-scale , which for massive stars, tends to be much shorter than the accretion time-scale (Hosokawa \& Omukai 2009).  

Rotation of protostars is likely regulated by magnetic fields generated by stellar dynamos.  In the absence of a magnetic field, accretion from the inner-edge of an accretion disk will tend to spin up the forming star to near brake-up speeds.   Convection in a spinning star drives dynamo action which amplifies magnetic fields.    Strong stellar fields which rotate rigidly with the star will extend beyond the stellar photosphere and interact with the inner portions of the disk.  The stellar rotation rate will be determined by magnetic disk braking, set by the radius and orbital period where the energy density of the field, $\rm B^2 / 8 \pi $, is comparable to the kinetic energy density of the disk, $\rho _{d} c^2_s(T_{d})$.    

Observations show that most  YSOs have spin periods of a few days.   This indicates  that for for Solar mass stars, stellar magnetic fields must reach out to a radius 2 to 10 stellar radii. The  disk interior to the critical radius where the stellar field is locked to the disk can be disrupted by the field which interacts with the charged component of the disk.   Ion-neutral coupling will drag the neutrals along with the magnetically-guided charges.   Instead of accreting directly onto the equator of the star, magnetic fields are thought to funnel accreting gas onto high-latitude regions of the YSO in so-called funnel flows.   If the dynamo-generated magnetic field strength is proportional to the stellar spin rate,  a balance will be reached when the spin of the star is just right.    Too slow a spin, the disk will reach the stellar surface and deliver high-angular momentum material to spin-up the star.   If the field is too strong, the field will reach to a larger disk radius where the orbital period is longer and slow down the stellar spin. 

\begin{figure}[!ht]
\begin{center} 
\includegraphics[width=6in]{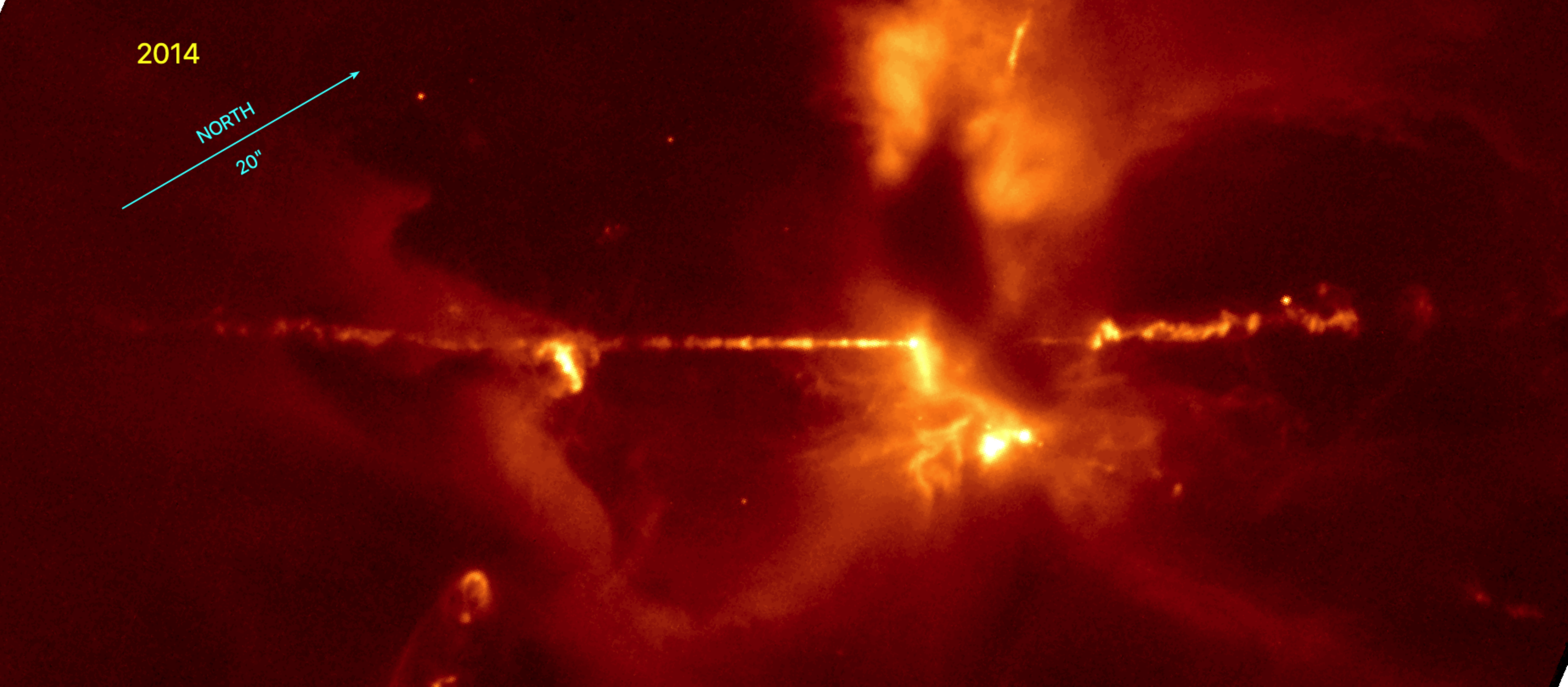}
\end{center}
\caption{An HST image showing 1.64~\mm\ [\Feii ] emission from  the HH~24 bipolar jet.  The jet is produced by a moderate-mass (2 to 3~\Msol ) YSO embedded in the dark cloud right of center.        This core contains a half-dozen YSOs.  Several other faint jets emerge from the region.  See  Reipurth et al. (2023) for details. }
\label{fig4}
\end{figure}

\section{Jets and Bipolar Outflows}

Outflows provide one of the easiest to observe indicators of the birth of a star.    Outflows can often be detected even when infrared emission from the forming star can not be seen (e.g. as in HH~211 -  Figure 1).
Magnetic fields can drive outflows during all phases of star formation (for a review, see Pudritz \& Ray 2019).  Torsional Alfv\'en waves during the  initial collapse of a spinning core can drive a slow bipolar flow along the spin axis.     Subsequently, as a
spinning protostar becomes fully convective, an alpha-omega dynamo can greatly amplify initial seed magnetic fields.   The dipolar component of these dynamo-generated fields can drive the so-called X-winds from the protostar.     While the resulting wind comes off the star as a wide-angle flow, hoop stress resulting from stellar (and disk) rotation can collimate this wind into a bipolar jet at a distance of 0.1 to 10s of AU from the  star
(Frank et al. 2014).    Figure 4 shows an example of a jet emerging from a forming, moderate-mass YSO.

Open field lines dragged in from the parent clump and core and connected to a spinning accretion disk  will develop a pinched, hourglass configuration with a helical twist symmetric about the rotation axis.  Ions and electrons produced by cosmic rays and radiation produced by accretion shocks  can be injected onto these open field lines. Centrifugal forces resulting from the pinched hourglass and twisted field geometry can accelerate charged particles to several times the orbital speed at the footprint of the field line in the disk.    The charged component will drag neutrals along by means of ion-neutral collisions to produce a so-called MHD disk wind.  The part of the MHD disk wind originating from the innermost portions of the disk will have the highest speed.  Parts of the wind originating at larger radii will have lower speeds.   As with X-winds, the azimuthal component of the twisted hourglass-geometry of the field lines results in hoop stress.  The hoop stress tends to focus the initially wide-angle (or isotropic) winds  launched from above and below the disk surface into a pair of oppositely directed jets at a distance of 1 to 100 AU from the star.  Figure 5 shows an illustration of the evolving magnetic field geometry of a collapsing and spinning core.

\begin{figure}[!ht]
\begin{center} 
\includegraphics[width=6in]{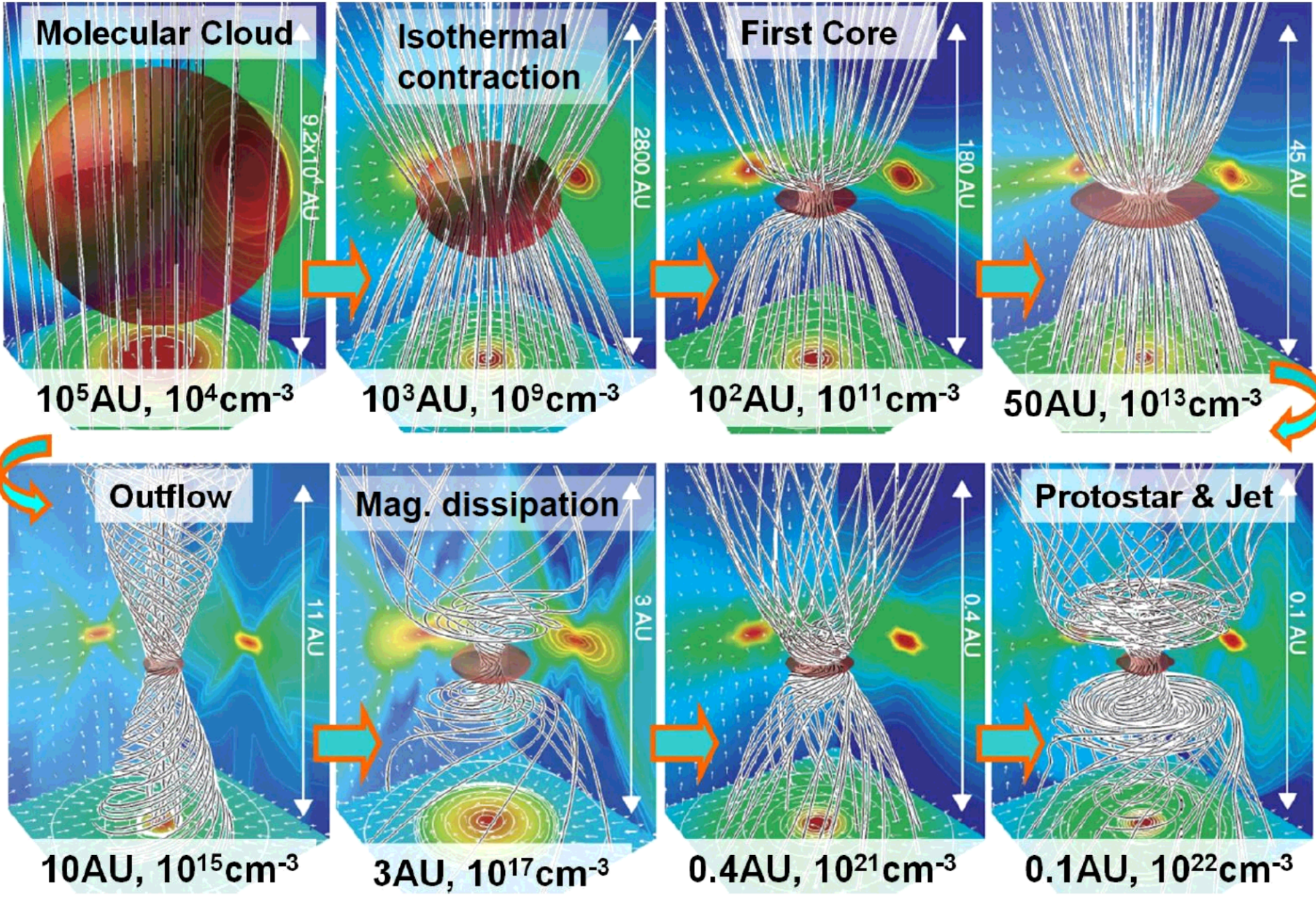}
\end{center}
\caption{A cartoon showing the evolution of the magnetic fields dragged into a protostellar disk by the gravitational collapse of a spinning core.  Note the pinched hourglass geometry.   The quasi-Keplerian rotation of the disk, combined with the wide-opening angle of the pinched field-lines can drive a magneto-centrifugal disk wind.   Additionally a dynamo-generated stellar field can power a faster  X-wind in the interior of the disk wind.   See Machida (2017) for details. }
\label{fig5}
\end{figure}

The structures and velocities of jets and outflow lobes provide  fossil records of the recent mass-loss, and through the intimate connection between accretion and mass-loss, the accretion histories of forming and young stars.   Measurements of the masses, momenta, and kinetic energies of outflows provide direct measures of the feedback impacts of protostars on their environments.    Visual and near-IR emission lines mostly trace  active shocks where fast ejecta runs into slower moving flow components (internal working surfaces), or where ejecta slams into  the ISM surrounding the forming star (terminal working surfaces).   Most shocks in star forming regions cool and stop radiating in years to decades.  Thus, HH objects tend to trace currently active shocks.    

The mm-wave emission lines of SiO trace material that has passed through shocks which liberated silicon from grains which subsequently interacted with oxygen.    This process can enhance the gas-phase SiO abundance in a molecular cloud by orders of magnitude, making this species an ideal tracer of recently formed shocks.  SiO persists in the gas phase for $\rm 10^2$ to over $\rm 10^3$ years.    CO emission is a long-term tracer of outflow activity since its low-lying mm-transitions are easily excited by collisions in 10 K gas.   CO emission will  be a good tracer of the outflow as long as the radial velocity of the swept-up gas can be distinguished from the host cloud.  Thus, CO is the best tool to estimate the total mass and momentum in the molecular component of an outflow.   However, many outflows blow out of their parent cloud cores or clumps and interact predominantly with ionized or atomic components of the ISM.   In this case, CO only provides a lower bound in the outflow momentum and kinetic energy.  Atomic hydrogen would be a potential tracer, but the 21-cm line from the outflow is very hard to distinguish from background emission from the Galaxy.
The HH~46/47 outflow (Figure 6) is a good example of a flow which has broken out of its parent cloud.   

\begin{figure}[!ht]
\begin{center} 
\includegraphics[width=6in]{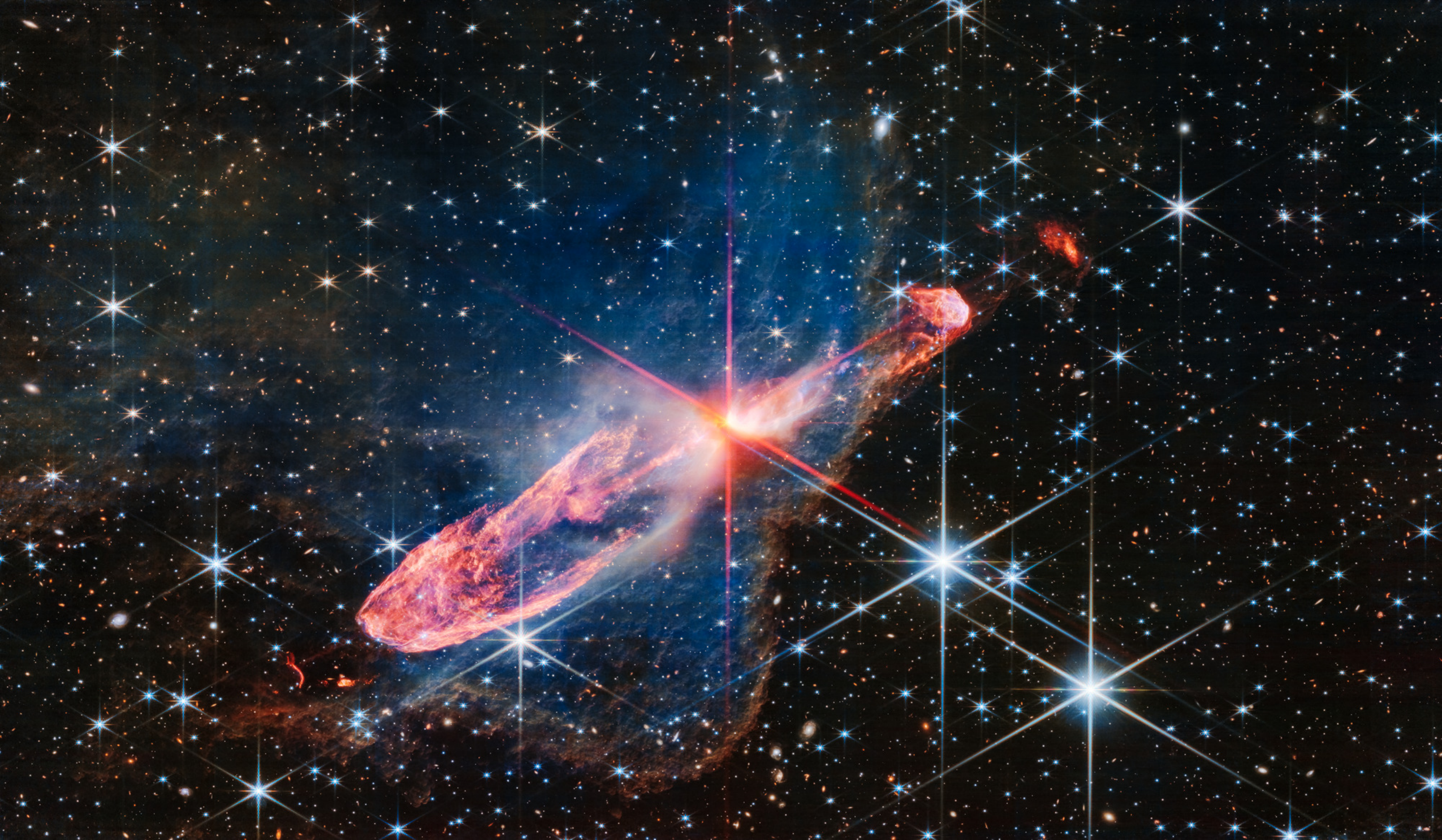}
\end{center}
\caption{A JWST image showing the HH~46/47 bipolar outflow from an isolated, small cloud embedded in the
Gum Nebula.   The source is located in the center and indicated by the red diffraction spikes.  Red colors trace shock-excited \Htwo\ emission.      The jet beam moving towards the upper-right blows out into the low-density medium surrounding the cloud.  Towards the lower-left, the beam burrows into the cloud where it inflates a shock-excited bubble.    North is towards the bottom-right.   The ionization front at the surface of the small cloud can be faintly seen below the outflow.  }
\label{fig6}
\end{figure}

The most commonly observed emission lines in HH objects are \Ha\ and  [\Sii ].    In photo-ionized plasma with Solar abundances, the intensity of the $\lambda \lambda$ 6717 / 6731 \AA\ [\Sii ] doublet has an intensity of about 0.1 times the intensity of \Ha .   However, in most shocks the intensity  of  the [\Sii ]  doublet is comparable to the intensity of  \Ha\ and can be used to distinguish shock-ionized from photo-ionized plasma.  These tracers probe ionized, or partially ionized gas with temperature of a few thousand K or warmer.  Other visual wavelength emission lines commonly seen in HH objects are [\Oi ], [\Oii ],  [\Oiii ], and [\Nii ].  In the near-IR, the 1.27 and 1.64 \mm\ lines of [\Feii ] are often bright.   When the jet and outflow is predominantly molecular, the near-IR emission lines of \Htwo\ are the best tracers of the shocks.  From the ground, the 2.12~\mm\ V=1-0 S(1) line is the most commonly observed transition.  Figure 7 shows a cluster of MHO outflows emerging from protostars embedded in the L1688 portion of the $\rho$-Ophiuchus cloud which is spawning a cluster of about 100 stars.

\begin{figure}[!ht]
\begin{center} 
\includegraphics[width=6in]{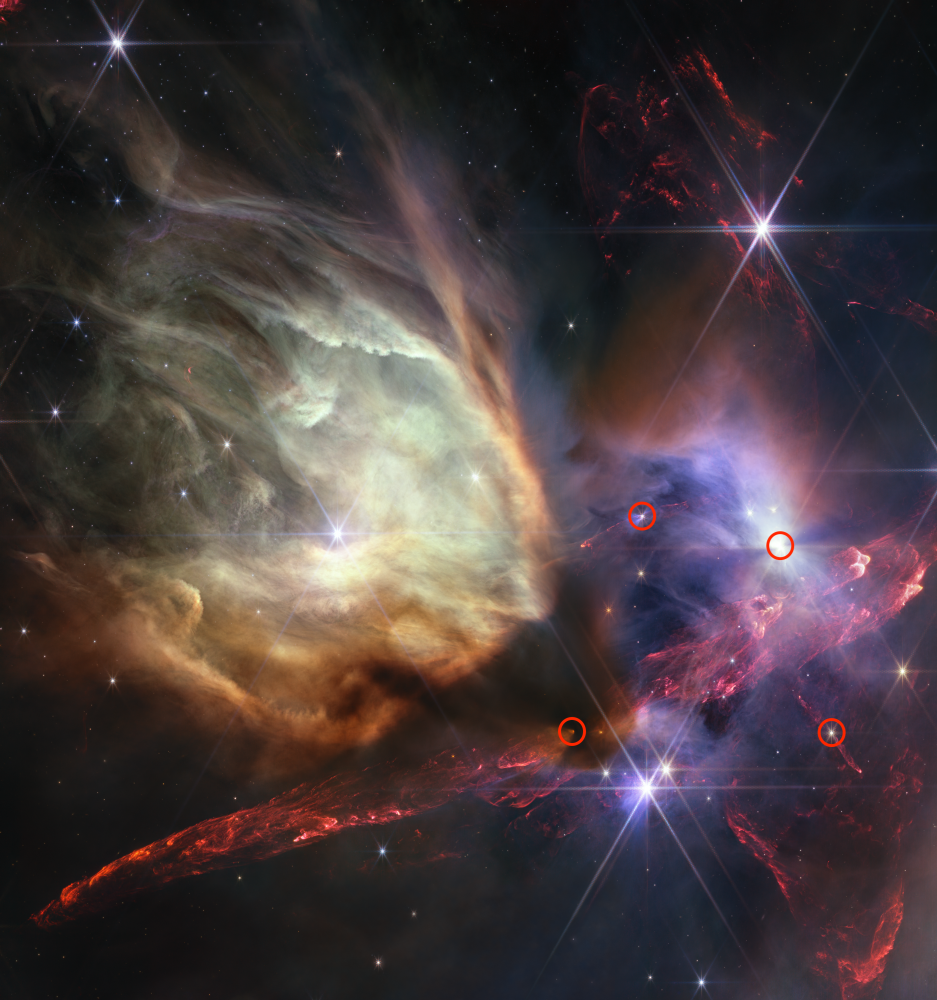}
\end{center}
\caption{A JWST image showing multiple outflows emerging from cloud cores in the Lynds 1688 portion of the $\rho$-Ophiuchus star forming region located  at a distance of $\sim$120 pc.     
Red circles mark four YSOs which drive outflows rendered visible in the 4.69~\mm\ \Htwo\ emission line.   The largest flow emerges from the highly embedded 
Class 0 protostar, VLA~1623.     The bright, conical, near-IR reflection nebula is illuminated by the moderately massive star GSS30. 
                  }
\label{fig7}
\end{figure}

Visual wavelength emission from most outflows requires the presence of active shocks.   However, in EUV-irradiated environments, the jet body and outflow lobes can be directly visible independent of shocks.
Irradiated outflows permit the measurement of flow properties using the standard theory of photo-ionized plasmas which can provide a more robust method of density measurement than the analysis of the highly non-linear theory of shocks. External radiation renders visible much weaker jets and outflows than the flows seen in dark clouds where only shock processed  gas can be seen at visual and near-IR wavelengths.  Figure 8 shows a pair of externally irradiated jets, HH~901 and HH~902,  emerging from YSOs embedded at the tips of photo-ablating pillars of molecular gas embedded within the Carina Nebula.

\begin{figure}[!ht]
\begin{center} 
\includegraphics[width=6in]{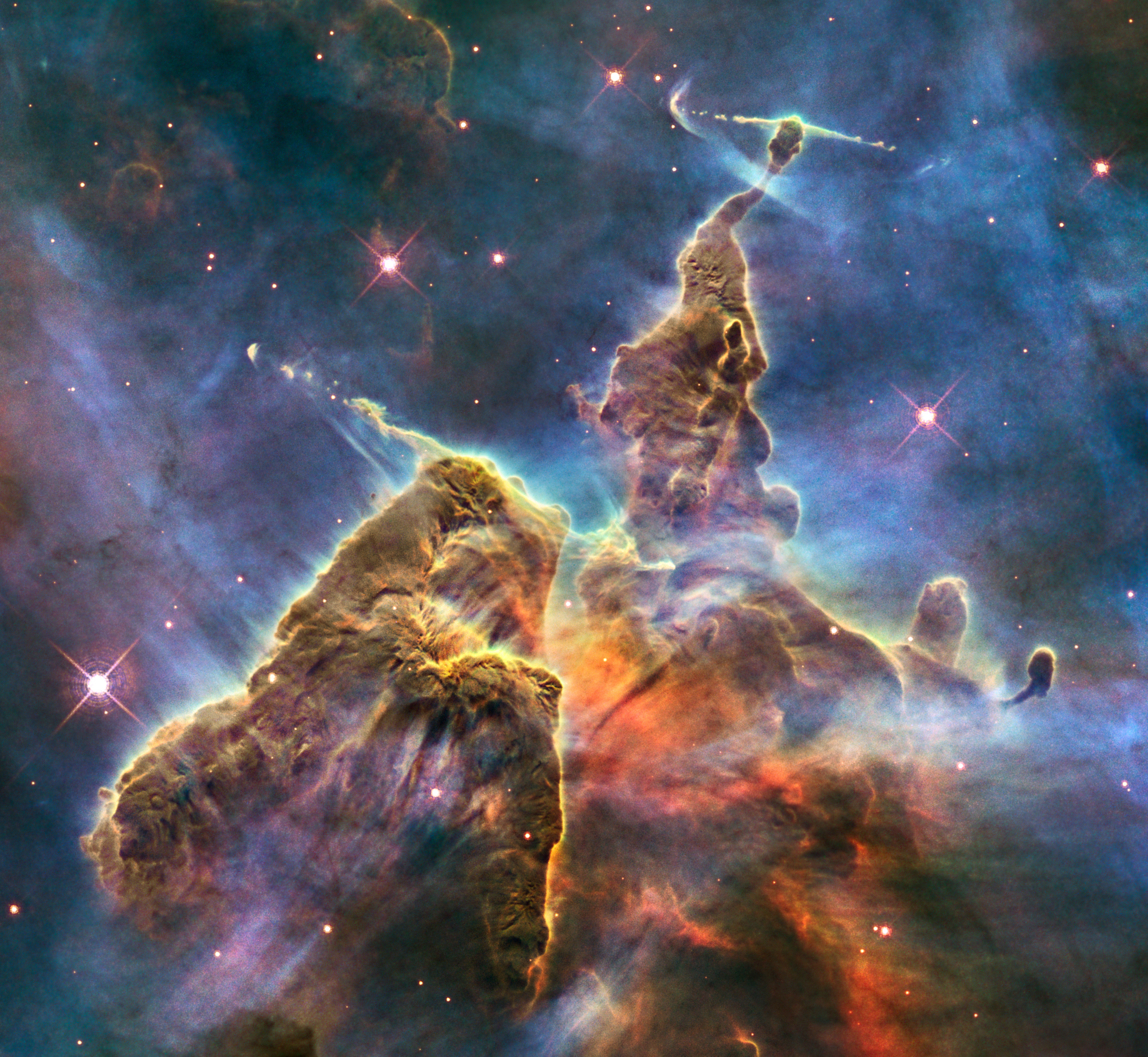}
\end{center}
\caption{An HST  image showing the HH~901 and 902 jets embedded in the Carina Nebula.  Intense EUV radiation illuminates the jet beams, allowing these irradiated jets to be seen even if no active shocks exist.}
\label{fig8}
\end{figure}

Jet symmetries provide clues about the nature and immediate environment of the source YSOs.   C-shaped bends indicate interaction of the jet with a side wind.   Alternatively, if the jet source is moving with respect to the surrounding ISM, the interaction of the jet with the medium can result in C-shaped deflections of the outflow.    Jets normally emerge along the disk axis.  S-shaped  bends can be an indication of forced precession of the accretion disk surrounding the protostar.      If the disk precesses,  the jet-axis will wobble and the ejecta will trace a point-symmetric helical pattern whose inner part exhibits a smooth S-shaped distribution of shocks.  Disk precession is usually a sign of a companion whose orbit plane in misaligned with the plane of symmetry of the disk around the jet source.   Abrupt disk orientation changes can be induced by companions in highly eccentric orbits, major accretion events from the cloud core or clump onto the disk, or the close passage of a star in the parent cluster.   Such events can results in Z-shaped outflow symmetry.  
 
Most HH jets have speeds between 100 to 300 \kms .  However, jet speeds as slow as 10s of \kms\ are seen for outflows from the youngest and least massive protostars.  Massive YSOs tend to have fast jets and more powerful outflows.   The $\sim$20~\Msol\ protostar IRAS~18162-2048 drives the fastest jet known.   Radio proper motions are as large as $\sim$1,200 \kms .  Some features in the HH~80 and  HH~81 shocks powered by this jet also have proper motions of $\sim$1,200 \kms\  (Bally \& Reipurth 2023).    When the proper motions are corrected for inclination angle, the  speeds of the fastest flow components in this outflow are close to 2,000~\kms .   HH~80 and HH~81  are bright in X-rays, indicating post shock temperatures of several million degrees.   Where this fast jet breaks out of the host cloud, it inflates a parsec-scale bubble of \Ha\ emission.  End to end, this outflow has a projected length of $\sim$10~pc.

\section{Protostellar Explosions}

\begin{figure}[!ht]
\begin{center} 
\includegraphics[width=6in]{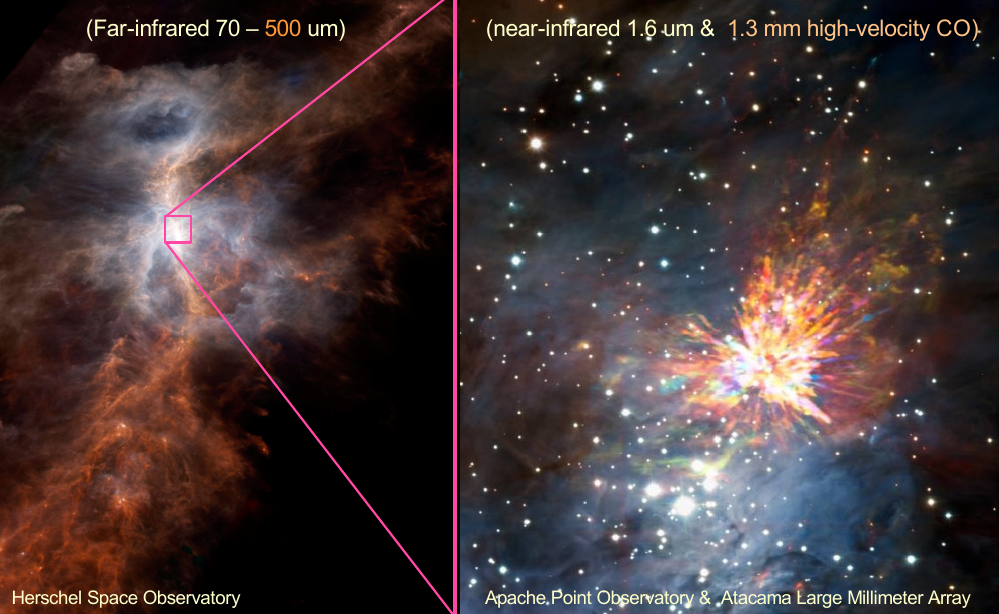}
\end{center}
\caption{ {\bf (Left:)} 
               The northern portions of the Orion A cloud known as the `Integral Shaped Filament'  in a Herschel Space Observatory 70 and 500~\mm\   image.  
              {\bf (Right:)}
              A near-IR mage showing the core of the Orion Nebula at 1.6~\mm .  The Trapezium stars responsible
              for ionizing the Nebula are seen near the bottom.  An ALMA  1.3 mm CO image showing  the molecular streamers associated with the OMC1 explosion in OMC1 is superimposed.                }
\label{fig9}
\end{figure}

During the last decade, it has become evident that in addition to jets and bipolar outflows, some massive protostars also produce powerful  but short-lived explosions at some point during their formation.  Orion Molecular Core 1 (OMC1), located about 0.1 pc behind the rear ionization front of the Orion Nebula,  contains the nearest group of massive protostars and one of the first discovered protostellar outflows (Figure 9).    Massive protostars in OMC1 include the $\sim$10~\Msol\ Becklin-Neugebauer (BN) object and $\sim$15~\Msol\ radio source I.   Radio proper motions have shown that these two objects are runaway stars ejected  in nearly opposite directions from the explosion center.   A third object, near-IR source x, a $\sim$3~\Msol\ star,  is also a runaway star.  

\begin{figure}[!ht]
\begin{center} 
\includegraphics[width=6in]{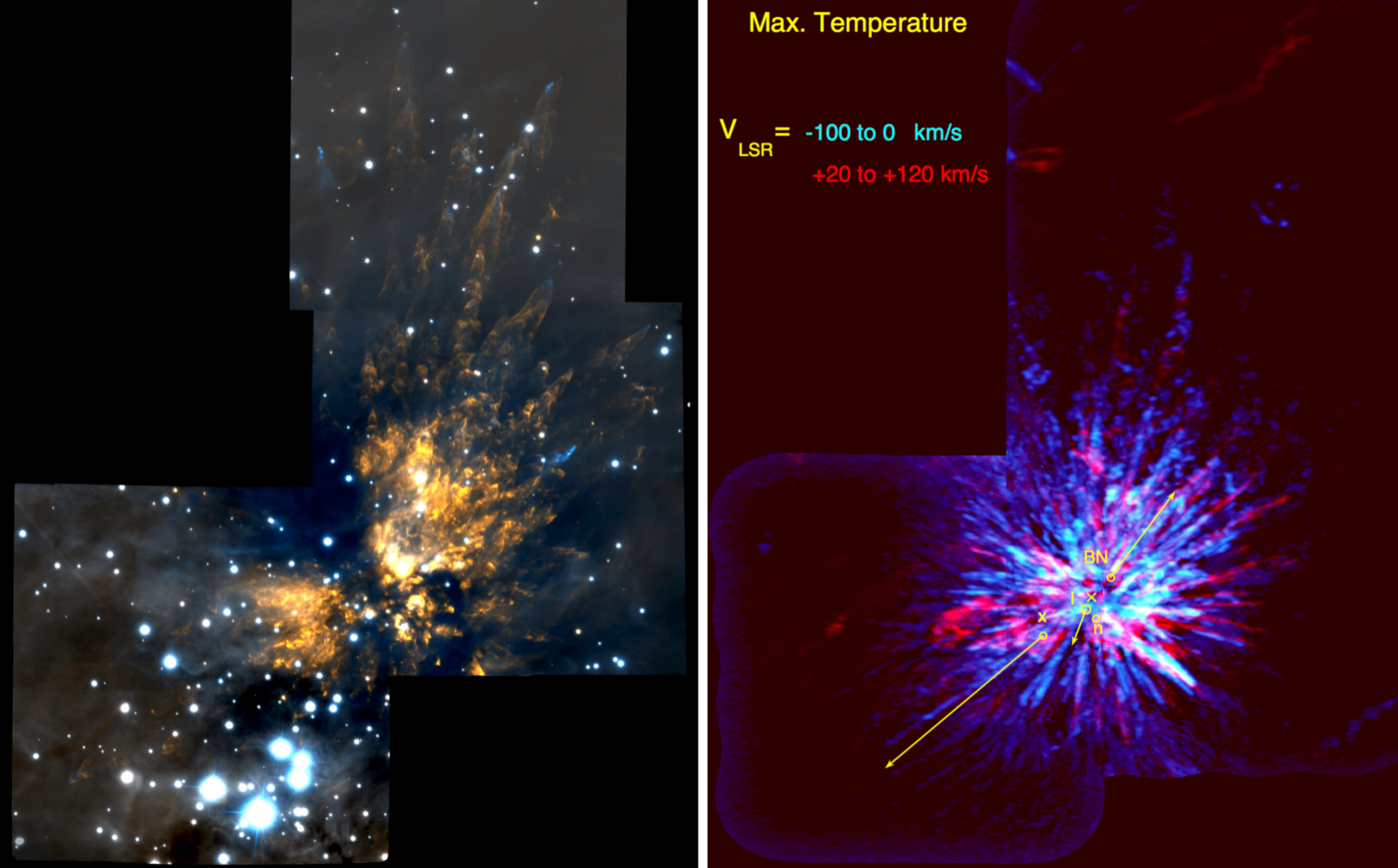}
\end{center}
\caption{ {\bf (Left:)} 
               The Orion explosion in the near-IR emission line of \Htwo\  at 2.12 \mm\ (orange)
               and the 1.644 \um\ emission line of [\Feii ] (cyan).
              {\bf (Right:)}
              ALMA image (at the same scale) showing 1.3 mm emission in carbon  monoxide (CO 2-1).
              Blue-shifted emission at radial velocities between V$_{LSR}$=-100 to 0 \kms\ with respect to
              the rest velocity of the Orion cloud, V$_{LSR}$=-9~\kms\ is shown in blue . 
              Red-shifted emission at radial velocities between V$_{LSR}$=+20 to +120 \kms\ is shown in red. 
              Small yellow circles show the locations of the $\sim$10~\Msol\ BN object, 
              $\sim$15~\Msol\  radio source I, and $\sim$3~\Msol\  source x.  These stars are moving away
              from the explosion center, marked by a yellow `x' symbol.    The yellow vectors show the
              expected motions of these three stars over the next 2,000 years.  The motions are based
              on the observed proper motions.
              Taken from Bally et al. (2020).
             }
\label{fig10}
\end{figure}

Near-IR images of shock-excited \Htwo\ have shown that the outflow from OMC1 consists of dozens of  shock-excited \Htwo\ fingers radiating away from a region between the BN object and radio source I (Figure 10, left).   Adaptive optics and ALMA images show that Orion OMC1 powered a $\sim 10^{48}$ erg explosion about 550 years ago (Zapata et al. 2009, 2013, 2017; Bally et al. 2020).  Figure 10 (right) shows CO streamers associated with the \Htwo\ fingers.    Most of the CO and \Htwo\ emission has radial velocities less than 50 \kms .   However, some CO emission extends to $\pm$120 \kms\ with respect to the host Orion A cloud.     The proper motions of the \Htwo\ fingers exhibits a Hubble flow character with the motions increasing linearly with projected distance from the explosion center.    The radial velocities of the CO streamers also exhibit a Hubble flow  (Figure 11).  The fastest proper motions are seen at the fingertips traced by  near-IR [\Feii ] emission.  Speeds reach $\sim$400 \kms\ at a projected distance of about 90 arcseconds from the explosion center.    Several of these finger-tips protrude into the Orion Nebula and emit in \Ha\ and [\Oiii ].

Figure 11 shows a spatial-velocity diagram showing the Hubble flow nature of the CO streamers in the OMC1 explosion.   The horizontal axis is a 90 arcsecond-long strip  extending from the south (left) to the north (right).    The vertical axis is radial velocity, extending from $\rm V_{LSR}$=-100 \kms\ at the bottom to +120~\kms\ at the top.   Three different north-south cuts through the data cube, separated by 1.5 arcseconds, are shown in red, green, and blue respectively.  The mostly resolved-out emission from the $\rm V_{LSR} \sim$9~\kms\ emission from the integral-shaped filament (ISF)  in this part of the Orion A cloud forms the horizontal band.
The roughly dozen streamers exhibit linear velocity gradients away from the explosion center.

\begin{figure}[!ht]
\begin{center} 
\includegraphics[width=6in]{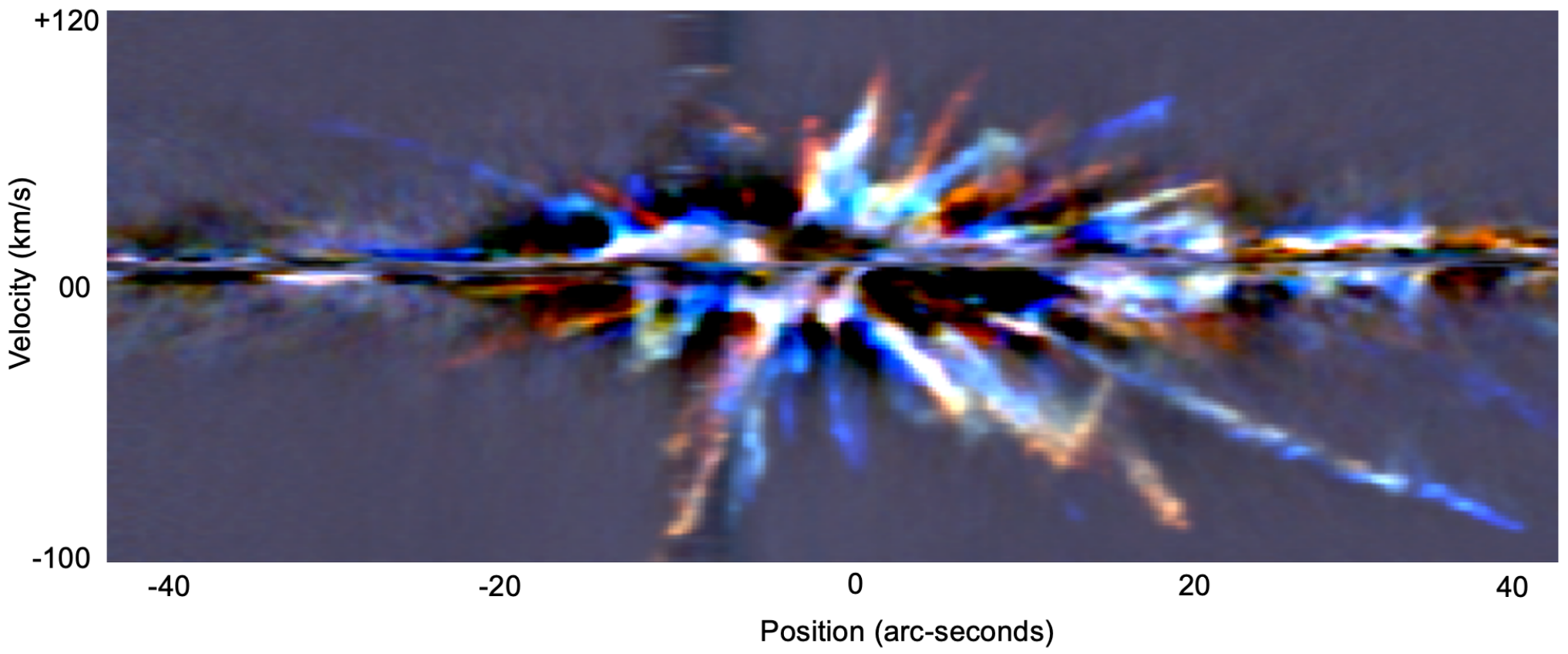}
\end{center}
\caption{ A north-south position-velocity diagram showing 1.3 mm CO 2-1 emission from some of the
molecular streamers the Orion explosion.   Three north-south strips separated by 1.5 arcseconds are shown
in red, green, and blue respectively.  The explosion center is near position 0.  The horizontal stripe above  V = 0 \kms\ shows the emission from the Orion A cloud ($\rm V_{LSR} \approx  $9~\kms ) which is mostly resolved out by the ALMA interferometer.    Note the `Hubble flow' nature of most streamers in which the radial velocity increased linearly with increasing projected distance from the explosion center.    This plot is effectively a high-resolution ($\sim$2 \kms ) 90 arcsewcond long, long-slit spectrum.              }
\label{fig11}
\end{figure}

The OMC1 explosion is associated with the ejection of three stars with speeds of 10 to 55 \kms .    The $\sim$10~\Msol\  BN object is moving towards the northwest with a speed of $\sim$30~\kms.  The $\sim$15~\Msol\ radio source I (its mass measured from the rotation curve of its nearly edge-on disk)  is moving towards the south with a speed of $\sim$10~\kms.  Finally,  $\sim$3~\Msol\  source x is moving towards the southeast with a speed of $\sim$55~\kms.   All three stars were located within 1 arcsecond of the explosion center about 550 years ago, implying that there is a direct connection between the dynamical interaction which ejected these three runaway stars and the explosion in the gas.   

The OMC1 core contains one of the densest known concentrations of YSOs with a mean projected separation of only about 2 arcseconds ($\sim$800 AU).   Such a high density increases the likelihood of interactions.  
The OMC1 explosion was likely caused by a dynamical interaction between a pair of forming binary stars embedded in the OMC1 core.  The stars were likely surrounded by disks.  Models show that in the interaction the binaries may have exchanged partners.  Two stars - the BN object and source x were ejected.  Two other stars formed either a compact binary or resulted in a protostellar merger.    The gravitational binding energy released by the formation of a more compact system powered the ejection of the stars and the explosion in the gas.   The sum of the kinetic energy of the ejected stars and the gas - $10^{48}$ erg - sets a constraint on the size-scale of the binary.   If the compact binary consists of a pair of equal mass stars whose total mass is $\sim$15~\Msol , then the mean separation of the stars has to be less than 2 AU.  If the stars have unequal masses, the separation must be less.   If the more massive member was accreting at a high rate and was bloated with an AU-scale photosphere, it likely would have interacted strongly with the lower-mass companion, possibly leading to engulfment resulting in a merger.   Source I in Orion may be either an AU-scale binary or a merger remnant.   

\begin{figure}[!ht]
\begin{center} 
\includegraphics[width=6in]{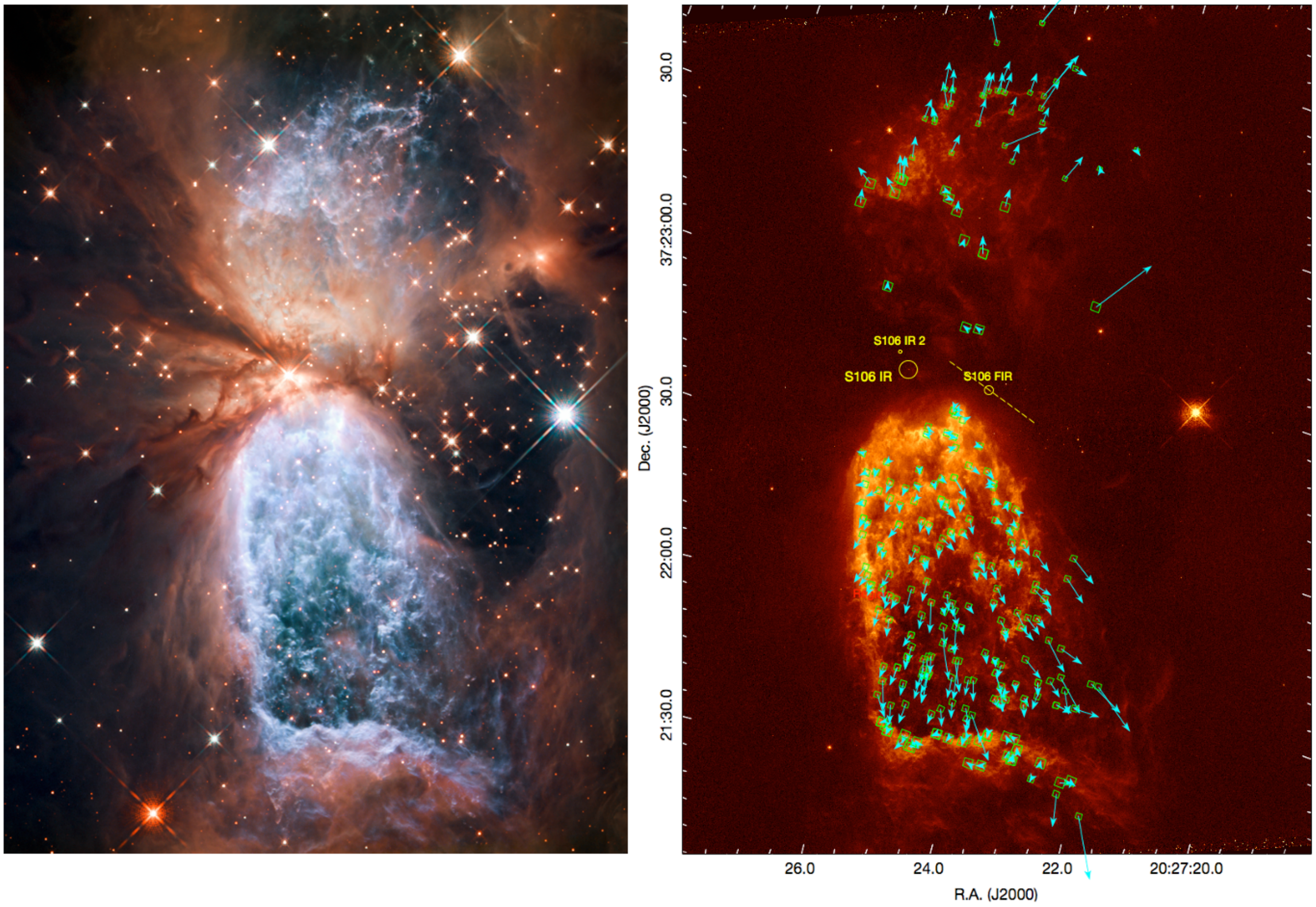}
\end{center}
\caption{
   {\bf (Left:)}  An HST image showing the very young, bipolar \Hii\ region 
   Sh2-106 (S106) in a color image.   
   Emission in the 1.6~\mm\ F160W filter is shown in red.  
   Emission in the 1.2~\mm\  F110W filter is shown in blue.  These images 
   were obtained in 2011 using HST/WFC3.     
   {\bf (Right:)}
   The 2011 epoch HST images showing [\Nii ].  
   Vectors show the highly supersonic, `Hubble flow' expansion of compact 
   knots in S106.  Proper motions were measured using \Ha\  images 
   obtained in 1995 together with the 2011 epoch HST images.
   The longest  vector corresponds to V$_{PM}$=170 \kms .
   The highly-obscured
   22 \Msol\ star which ionizes the nebula, S106-IR, is marked.  
   S106~IR~2 is a low-mass sub-mm core.  S106~FIR is a Class 0 protostar 
   driving an outflow whose direction is indicated by the dashed yellow line. 
   Taken from Bally et al. (2022). }
\label{fig12}
\end{figure}

The very young bipolar \Hii\ region Sh2-106 (S106)  exhibits signs of an explosion about 3,500 years ago (Figure 12).  At a distance of 1.09 kpc, it is the second nearest protostellar explosion, next to Orion OMC1.   Its highly obscured ($A_V \sim$20 magnitudes) ionizing central star is embedded in a nearly-edge-on disk which gives S106 its unique bipolar shape.  Multi-epoch HST images show that compact knots in the nebular lobes have proper motions increasing linearly with projected distance from the massive protostar, S106-IR (Bally et al. 2022).    Protostellar explosions have also been found in  G5.89,   DR21  and IRAS~12326-6245 (Zapata et al. 2009, 2013, 2017, 2020, 2023).    Explosive outflows are a particularly important feedback channel in the self-regulation of star formation because they inject momentum and energy directly into their parent clouds. 

The half dozen known protostellar explosions associated with massive star forming regions implies that the Galactic event rate must be about one per century,  comparable to the supernova rate.   Explosive signatures are short lived because these are impulsive events.   If the time to form a massive star is $\sim 10^5$~years, and the typical survival time of explosive signatures is a few thousand years, one would expect that only a few percent of massive star forming regions exhibit explosive signatures at any one time.

\section{Conclusions}

Stars form from the gravitational collapse and fragmentation of molecular clouds (GMCs) which
produce clumps and star-forming cores.   The duration of star formation in a typical GMC such 
as the clouds in the Orion region is about 10 to 20 million years. Most star formation occurs in clusters.
But the low efficiency of star formation (5 to 30\%) in clumps and about 5\% for an entire GMC, implies 
that most young clusters are short lived.   As low-mass stars evolve onto the main sequence (in $\sim$30
Myr for Sun-like stars) the massive stars formed in the same star-forming regions are dying.  Their
UV, winds,  and supernovae destroy the parent cloud, create \Hii\ regions, and superbubbles.    Atoms
in CMCs have about a $\sim$5\%  chance to end up in a star during  $\sim$100 Myr Galactic ecology cycle as
GMCs are destroyed, converted into the hotter ionized and atomic phases of the ISM, and re-condensed into
new GMCs.  
 
In a collapsing star-forming core, entrained magnetic fields are amplified by compression, shear-dynamos in disks, 
and convective-dynamos in the forming protostars.  Magnetic fields are responsible for 
regulating stellar rotation, for removing angular momentum from disks to allow accretion onto
growing stars, and for launching outflows and collimating these flows into jets.  Outflows 
can be launched by the protostar itself in a so-called X-wind, or by the disk in a disk wind.
  
Accretion and outflow are intimately connected.   Variations in accretion rate result in variations in mass-loss
rate and outflow ejection velocity.   These variations are responsible for the complex structure and
kinematics of protostellar outflows.   As these primary winds and jets interact with the surrounding ISM, they
sweep up bipolar shells.  As flows develop in a GMC, they create bipolar molecular outflows.  When
they burst out of their host molecular clouds, they sweep up shells of atoms or  plasma.
Outflow symmetries and structure provide fossil records of the mass-loss and accretion histories 
of the source young stellar objects.     Outflow power increases with source mass and luminosity.  

In addition to bipolar outflows,
some massive star forming regions such as OMC1 produce powerful explosions.    Explosions appear to be powered
by N-body interactions and the decay of non-hierarchical multiple systems as they re-configure into 
more compact, hierarchical configurations and ejected stars.  The currently known half-dozen explosive protostellar
outflows in our region of the Galaxy implies that the event rate is comparable to the Galactic supernova rate.  
Thus, most massive stars may experience an explosion at some point during their formation.  Because explosions are impulsive, their signatures are short lived compared to the time required to form a massive star.   Thus, at any one time,
most forming massive stars are likely  to be associated with bipolar flows.     
Only a few percent are likely to exhibit explosive signatures.

\section{Acknowledgements}

The work presented here was supported by the National Science Foundation in the US  by NSF grant AST-1910393.

\end{document}